# Correction of Variation due to Non-Hydrostatic Effects The Observed Temperature in Upper-Air Sounding.

Kochin A. V.

Central aerological observatory, Roshydromet,
Pervomaiskaya st. 3, Dolgoprudny, Moscow Region, Russia
Email: *amarl@mail.ru*

*Abstract*

*There are a lot of oscillatory motions of various kinds in the atmosphere, for example, internal gravity waves (IGW), which have a period less than the time of flight of the radiosonde. Oscillatory motions lead to adiabatic cooling during the shift of air mass upwards from equilibrium or heat when shifting down. The magnitude and sign of the shift from the equilibrium state of a specific volume of air at the time of measurements by radiosonde are unknown. Therefore the displacement of measured (essentially instantaneous) temperature is unknown relative to the equilibrium temperature. These errors in some cases can reach values of 10 degrees or more. The algorithm for correcting of these effects is proposed. The algorithm of correction calculates the shift of the vertical displacement of the radiosonde from the polynomial curve. The correction for temperature is calculated as a product of the dry adiabatic lapse rate γa (-0,01 °/m) and the shift of the radiosonde altitude. The result is subtracted from the measured temperature. The observed height is corrected in dependence of the shift value.*

**1. Introduction**

The vertical profile of meteorological parameters (temperature, pressure) according to the radiosonde is calculated in the hydrostatic approximation, i.e. the assumption that the vertical profile of air temperature does not change during the flight of the radiosonde. In this approximation, the meteorological parameters are then used in different applications (weather forecast, etc.). However there are oscillatory movements of various kinds in the atmosphere, for example, internal gravity waves (IGW), which have a period much less than the time of flight of the radiosonde. The vibration motions lead to adiabatic cooling when the displacement of the air mass upwards from equilibrium or heat when shifting down. These errors can be characterized as errors of representativeness. The magnitude and sign of the displacement from the equilibrium state of a specific volume of air in the time of flight of a radiosonde it is unknown, therefore it is also unknown and the offset measured (in fact instantaneous) temperature relative to the equilibrium temperature. The errors can reach a value of 10 degrees or more. This paper analyzes the mechanism of occurrence of temperature drift due to the oscillatory motion of the air and proposed method of correction of measured temperature values.

**2. The mechanism of occurrence of errors in temperature measurement due to the fluctuations of the atmosphere**

The fluctuations in the atmosphere are permanent and are recorded in almost any weather conditions. The process of formation of oscillations in the atmosphere has been studied for a long time, the results of research lit, for example, in the monograph E. Gossard, W. Hooke (1). The internal gravity waves (IGW) give the greatest influence on the temperature sensor of radiosonde. IGW have a period in the range of 200 to 1000 seconds approximately. The period of oscillation of IGW is determined by the equation of Brunt-Väisälä. IGW have a significant vertical component, because returning force for IGV is the force of buoyancy which directed vertically. Consider the mechanism of the displacement measured by the radiosonde temperature from the equilibrium in a dedicated volume of air. Assume that the air in a selected volume of air is stationary. Then the temperature of air in the volume does not depend on time and equal to the equilibrium temperature T. If the air in the selected volume is periodically shifted up or down, the temperature will be equal to the equilibrium only when the volume is in the middle position. If the displacement volume is shifted up, the temperature will be less due to the adiabatic expansion, when shifting down, the temperature will become higher due to compression. If we could measure the temperature over time more of the oscillation period and averaging the results, we will obtain the original equilibrium temperature T.

The process of measuring temperature using radiosonde is as follows. A radiosonde that is attached to the balloon, filled with hydrogen or helium, moving with constant speed relative to the ambient air, which is determined by the balance between aerodynamic drag force and lifting force. The temperature sensor of the radiosonde at some intervals of time measures the instantaneous air temperature and telemetry channel transmits the results of to a ground station. If fluctuations is absent in the atmosphere, and the temperature is constant, then the sensor will give constant temperature. If the volume of air is displaced periodically up or down, and the temperature will also be displaced from the equilibrium temperature T. Accordingly, the vertical temperature profile will be a kind of undulating curve, although in reality, the equilibrium temperature T does not change. To obtain an equilibrium temperature dependence of the some height, the radiosonde would have to stop at each of the heights for a time greater than the oscillation period, to measure a series of values and averaging them. Then the result would correspond to the equilibrium temperature. The changing the speed of rising of the radiosonde indicates to the presence of errors due to the vertical air movements. If the radiosonde rises with a constant velocity, they measured temperature corresponds to the equilibrium. If its rate of rising changes, the measured temperature will have some error.

### 3. The actually observed temperature shifting

The reason for this investigation was the results of monitoring the quality of radio sounding of atmosphere conducted by Central Aerological Observatory. It was identified in the analysis of data the shifts at 10 degrees and more are very rare. It is usually less one degree. However, software that filters the data for the forecasting centers does not pass such cases. The author does not have statistics about the frequency of such cases. However, this needs to be corrected even if these are isolated cases per a year.

### 4. The algorithm and results of correction

The testing of the correction algorithm lasted a very long time. The values of the vertical velocity were used in the first stage for temperature correction (Fig.1). However, attempt to develop this algorithm was not successful. Attempts to take into account a variety of effects, including Bernoulli's equation and the variability of the horizontal velocity, led to the fact that in some cases, the result was very good, but in other cases there was too large value of correction of temperature. The algorithm for the velocity proved to be unstable. The results of correction are on Fig.2. The maximum correction is 8 degrees.

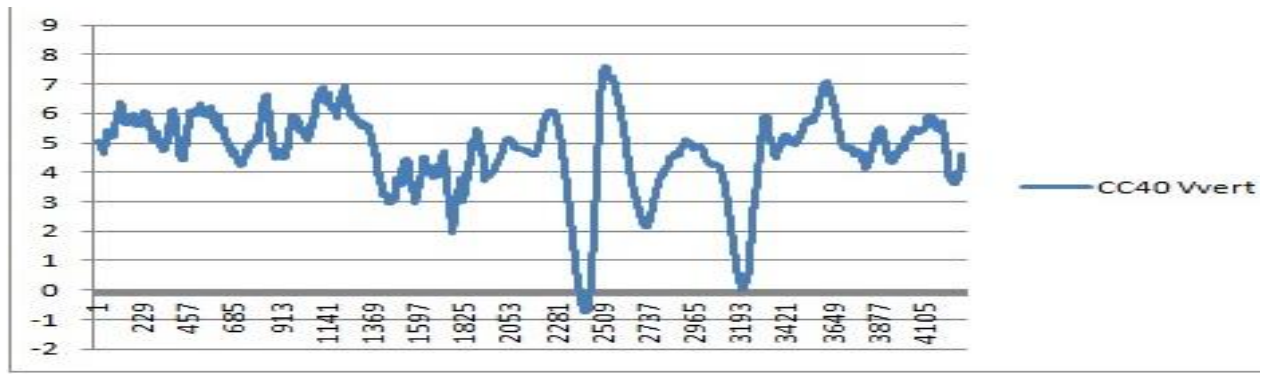

Fig.1. The vertical velocity of radiosonde. The y-ordinate is vertical velocity m/s, the x-coordinate is the time of the launch.

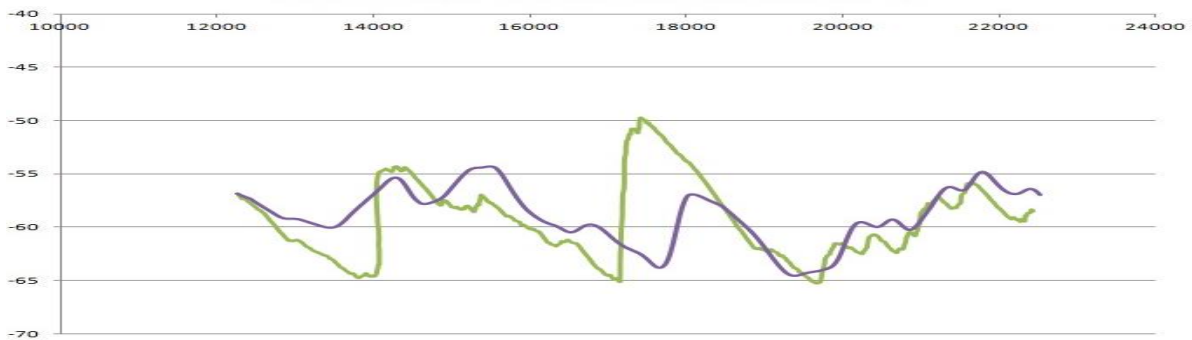

Fig.2. The results of correction. The green line is a raw temperature; the purple line is corrected temperature. The y-ordinate is temperature; the x-coordinate is the height in meters.

Then the algorithm was simplified and used only the values of the height of the radiosonde. The algorithm of correction calculates the shift of the vertical displacement of the radiosonde from the polynomial curve. The correction for temperature is calculated as a product of the dry adiabatic lapse rate $\gamma_a$ (-0,01 °/m) and the shift of the radiosonde altitude. The result is subtracted from the measured temperature. The observed height is corrected in dependence of the shift value. The results of correction are on Fig.3. The maximum correction is 5 degrees.

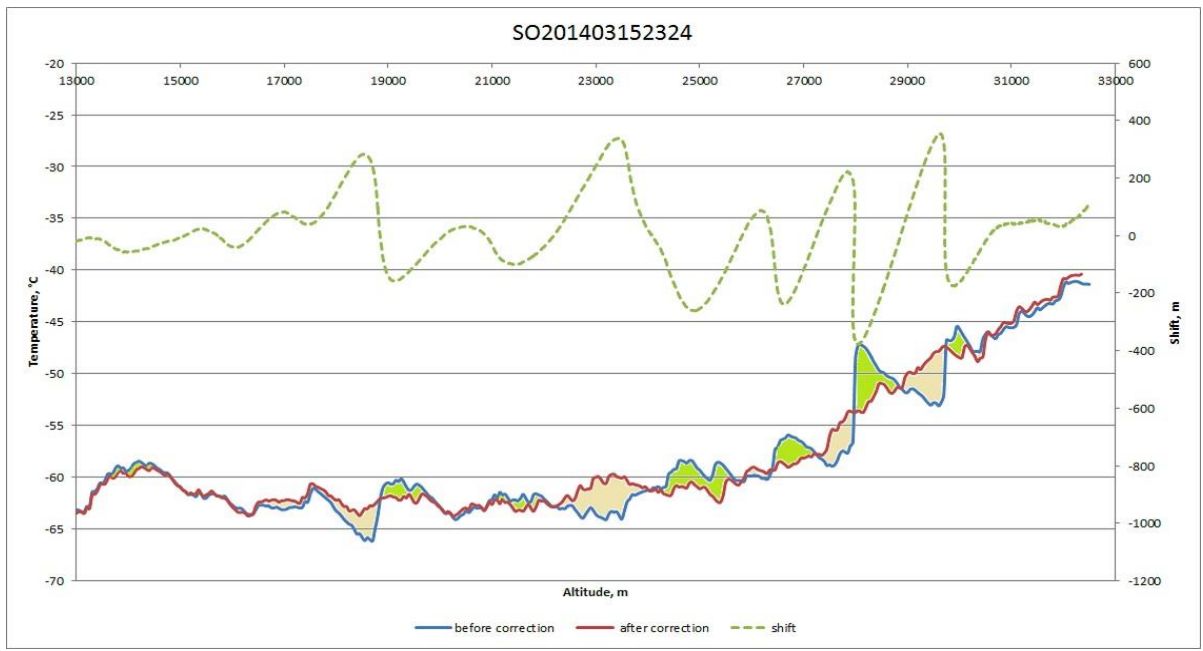

Fig.3. The results of correction of the observed temperature in Sochi 15 March 2014. The green dotted line is the shift of the vertical displacement of the radiosonde from the polynomial curve. The blue line is the observed temperature. The red line is the corrected temperature. The areas where the air is heated due to the shift down from the equilibrium state are highlighted in green. The areas where the air is cooled due to the rise from the equilibrium state are highlighted in pink. The y-ordinate is temperature and vertical shift, the x-coordinate is the height in meters.

## 5. Conclusions

A method of correcting measurement temperature in the atmosphere taking into account nonhydrostatic effects arising from the appearance in the atmosphere of oscillatory movements is developed. It would be useful to hold the testing this method on the data sets in different climatic zones. In case of successful testing should develop software to implement the method in the process of upper air sounding

### ACKNOWLEDGEMENT

The author would like to thank colleagues, A. Kats especially, for discussion and A. Naumov for providing data and V.Fomenko and I. Gubarchuk for data processing.

### REFERENCES

*1.E. Gossard, W. Hooke, Waves in the Atmosphere. 1975.*